\documentclass[
journal=jacsat, 
manuscript=article]{achemso}

\usepackage[version=3]{mhchem} 
\usepackage{pgfplots}
\usetikzlibrary{positioning}
\makeatletter
\def\acs@author@fnsymbol#1{}
\makeatother

\author{V.Sai sumith reddy}

\email{saisumithreddy23@gmail.com}

\title[\texttt{achemso} demonstration]
{Particle motion and scattering in Finslerian Schwarzschild metric}

\begin{document}

\begin{abstract}
Finsler geometry is just riemannian geometry without the quadratic restriction[1]. In this paper, we study the motion of massive(non-zero rest mass) and massless particles for schwarzschild metric in finsler spacetime in the case of two dimensional Randers-Finsler space with unit positive flag curvature instead of two dimensional riemann sphere. We provide qualitative study of potential energy of particles at various distances near schwarzschild black hole and the results are then compared with schwarzschild metric in riemannian geometry. We will also describe the scattering of particles near schwarzschild black hole in finsler spacetime by finding capture cross section of absorption by the black hole.
\end{abstract}

\section{I.Introduction}

In  general relativity, gravitational force is due to distortion of matter with spacetime. General relativity relates curvature of spacetime to energy and momentum of matter and radiation. The classical tests of general theory of relativity are the perihelion precession of mercury's orbit, the deflection of light by the sun and gravitational redshift of light. Schwarzschild derived vacuum solutions for Einstein field equations under symmetry assumptions. The schwarzschild metric describes gravitational field outside a spherical body(of mass M,say) with zero electric charge and zero angular momentum. It has singularities at r=0 and r= 2M [7]. But r=2M is a coordinate singularity which could be removed by transforming the metric to retarded or advanced time coordinate.

Finsler geometry provides metric generalization to riemannian geometry which includes lorentz metric as a special case. Riemannian geometry defines inner product structure over tangent bundle where  as finsler geometry generalizes metric geometry by defining a general length functional for curves on the manifold[11]. By imposing few requirements on finsler structure, one could model a geometric spacetime which provides physical effects that are useful and suitable for physics[10]. Finsler geometry is used in various fields of research such as medical imaging, optics, relativity and cosmology[15]. 

Finsler spacetimes are generalizations of lorentzian metric manifolds which preserve causality. Xin Li et.al [11] studied exact solution of vacuum field equations in finsler spacetime and described it to be same as vanishing ricci scalar which implies that the geodesic rays are parallel to each other. Silagadze[18] proposed finslerian extension of the schwarzschild metric. Chang [5-6] argued that finsler geometry, in principle, can address the rotational curves of galaxies and the acceleration of the universe without introducing dark matter or dark energy. Minguzzi[14] studied singularity theorems, raychaudhuri equation and its consequences for chronality in the context of Finsler spacetimes. Recently, Fuster[15] presented finslerian version of the well-known pp-waves which generalizes the very special relativity (VSR) line element. Pfeifer [2] gave precise geometric definition of observers and their measurements, and showed that two different observers are related by a transformation composed out of a certain parallel transport and a lorentz transformation. In [10], Javaloyes described few links between finsler Geometry and the geometry of spacetimes. 

In section 2, we discuss the basic concepts and definitions of finsler geometry. In section 3, we will study motion of particle in schwarzschild metric and compare it with riemannian case. In section 4, we will study  the scattering of massive and massless particles. In section 5, we draw conclusions.

\section{II. Finsler Geometry}

Finsler geometry uses generalization of the notion of distance between two neighbouring points $x^i$, $x^i + dx^i$ which is given by a function $F(x^i,dx^i)$ with quasimetric structure and minkowski norm[17].\begin{equation} ds = F(x^i,dx^i) \end{equation}
\textbf{Quasimetric}: A metric d which satisfies the following three axioms is called quasimetric.

(i)Positivity : $d(x, y) \geq 0 $

(ii)Positive Definiteness : $d(x, y) = 0$  if and only if   $x = y$

(iii)Triangle Inequality : $d(x, z) \leq d(x, y) + d(y, z)$.

We can see that quasimetric need not satisfy symmetry.

\textbf{Note}: The derivative of quasimetric is given as 
\begin{equation}  d'(x, y) = \frac{1}{2}(d(x, y) + d(y, x)) \end{equation} 
\textbf{Minkowski Norm} : A function $F:TM\rightarrow [ 0,\infty )$ is a minkowski norm if $F$ is smooth, homogeneous of degree 1 and is of symmetric bilinear form[16].
\hfill \break
The finite distance between two points on a given curve $x^i=x^i(s)$ between $t_1$ and $t_2$ is given by the length functional[17] :
\begin{equation}  s = \int_{t_1}^{t_2} F(x^i(t),\frac{dx^i(t)}{dt}) dt \end{equation} 
\hspace{1cm} For s to be independent of parameter t, $F$ must satisfy homogeneity of degree one
i.e.,$F(x^i,k dx^i) = k F(x^i,dx^i)$.
\hfill \break
Consider an n-dimensional smooth manifold $M$. Let $T_xM$ be tangent space at the point $x \in M$. The set of all tangent spaces TM = $\{ T_xM  |   x \in M\}$ is called tangent bundle on the manifold $M$.

\hspace{1cm}We denote $(x,y) \in TM$ where $x \in M$ and $y \in T_xM$. 
\hfill \break
\textbf{Finsler Structure} : It is a function $F:TM\rightarrow [ 0,\infty )$ defined on a tangent bundle $TM$ which satisfy the following conditions[15]:

(i) Regularity : F is smooth on $TM $\textbackslash $\{0\} $.

(ii) Homogeneity : $F(x,ky) = k F(x,y)$ for all  $\lambda > 0$. 

(iii) Strong Convexity : The fundamental metric tensor 
\begin{equation}  g_{ij}(x,y) = \frac{1}{2}\frac{\partial^2F^2(x,y)}{\partial y^i \partial y^j} \end{equation} \hspace{1cm} with i, j = 1, . . . , n, is positive definite.

Note that the metric tensor $g_{ij}$ is symmetric under exchange of  indices.
\hfill \break
\textbf{Finsler Manifold} : A smooth manifold M equipped with finsler structure F is called finsler manifold (M,F).

\textbf{Note} : For pseudo-finsler spaces, metric tensor $g_{ij}$ need not obey positive definiteness[15].

Because of homogeneity, we get $F^2(x^k, y^k) = g_{ij}(x^k, y^k) y^iy^j$.
Hence $ds^2 =  g_{ij}(x^k, y^k) y^iy^j $.

The  space is riemannian if $g_{ij}$ does not depend on $y^k$ i.e., it depends only on position $x^k$ but not on velocity $y^k$.

The geodesic equation in finsler manifold is given by 
\begin{equation}  \frac{d^2x^i}{ds^2} + 2G^i = 0\end{equation} 
where s is parameterizing the geodesic curve.

The geodesic spray coefficients $G^i$ are given[11] by
\begin{equation} G^i = \frac{1}{4}g^{ij}(\frac{\partial^2F^2}{\partial x^k \partial y^j} y^k - \frac{\partial F^2}{\partial x^j}) \end{equation} 

The Finslerian Ricci scalar[15] is given as trace of tensor $R^i_k$ : $Ric := R_i^i$.
The tensor $R^i_k$ is defined[15] as :\begin{equation} R^i_k = \frac{2}{F^2}\left(\partial_{x^k} G^i - \frac{1}{2}\partial_{y^k}G^j\partial_{y^j}G^i - \frac{1}{2}y^j\partial_{x^j}\partial_{y^k}G^i + G^j\partial_{y^j}\partial_{y^k} G^i\right) \end{equation}
Here $\partial_{x^k} = \frac{\partial}{\partial x^k}$, $\partial_{y^k} = \frac{\partial}{\partial y^k}$.
The Finslerian Ricci tensor is given as :
\begin{equation}  Ric_{ij} = \frac{1}{2}\partial_{y^i}\partial_{y^j}(F^2Ric) \end{equation} 
It reduces to Ricci tensor when finsler structure F is Riemannian.
\hfill \break
\textbf{Berwald spaces} : If geodesic spray coefficients for a finsler space are quadratic in y, then such spaces are called berwald spaces[15].

Berwald spaces satisfy $Ric_{ij} = 0$.

\section{III. Particle motion in Schwarzschild metric in Finsler setting }
The finsler structure in Schwarzschild metric is given[11] as:
\begin{equation} F^2 = A(r)dt^2 - B(r)dr^2 - r^2\Bar{F}^2(\theta,\phi,d\theta,d\phi)\end{equation} 
Here A(r) = $(1 - \frac{2GM}{\lambda r})$,$B(r) = ( \lambda - \frac{2GM}{r} )^{-1}$.

The modified schwarzschild radius is given[11] as $r_s = \frac{2GM}{\lambda}$.

In the paper[19], Bao et.al gave a complete classification of Randers-Finsler space with constant flag curvature. For $ \Bar{F}^2$, we consider a two dimensional Randers-Finsler space with constant positive
flag curvature $\lambda = 1$ which is given by:
\begin{equation} \Bar F = \frac{\sqrt{ \left(1-\epsilon ^2 \sin^2 \theta \right)\text{d$\theta $}^2 + \sin^2\theta \text{d$\phi $}^2 }}{1-\epsilon ^2 \sin^2\theta}-\frac{ \text{$\epsilon $} sin^2\theta \text{d$\phi $}}{1-\epsilon ^2 \sin^2\theta }\end{equation}
Here $ 0 \leq \epsilon < 1$.
\hfill \break
The exterior metric of vacuum field equations is given[15] by :
\begin{equation} ds^2 = A(r)dt^2 - B(r)dr^2 -r^2\left(\frac{\sqrt{ \left(1-\epsilon ^2 \sin^2 \theta \right)\text{d$\theta $}^2 + \sin^2\theta \text{d$\phi $}^2 }}{1-\epsilon ^2 \sin^2\theta}-\frac{ \text{$\epsilon $} sin^2\theta \text{d$\phi $}}{1-\epsilon ^2 \sin^2\theta }\right)^2\end{equation} 

We use advanced time coordinates $(v,r,\theta,\phi)$ as 't' breaks down near the horizon and also radially infalling particle moves along line of advanced time.

So,the schwarzschild metric in $(v,r,\theta,\phi)$ coordinates is given as :
\begin{equation} ds^2 = V(r)dr^2 - 2dvdr -r^2\left(\frac{\sqrt{ \left(1-\epsilon ^2 \sin^2 \theta \right)\text{d$\theta $}^2 + \sin^2\theta \text{d$\phi $}^2 }}{1-\epsilon ^2 \sin^2\theta}-\frac{ \text{$\epsilon $} sin^2\theta \text{d$\phi $}}{1-\epsilon ^2 \sin^2\theta }\right)^2\end{equation} 
\hspace{1cm}Here $V(r) = 1 - \frac{2M}{r}$.

Consider an action
\begin{equation} I = \int g_{ij} \dot x^i \dot x^j ds \end{equation}
\begin{equation}I = \int \left(V(r)\dot v^2 - 2\dot v\dot r - r^2\left(\frac{\sqrt{ \left(1-\epsilon ^2 \sin^2 \theta \right)\text{$\dot \theta $}^2 + \sin^2\theta \text{$\dot \phi $}^2 }}{1-\epsilon ^2 \sin^2\theta}-\frac{ \text{$\epsilon $} sin^2\theta \text{$\dot \phi $}}{1-\epsilon ^2 \sin^2\theta }\right)^2\right) ds\end{equation}
\hspace{1cm}Here $. = \frac{d}{ds}$ , 's' represents propagation along worldline of the particle.

The extremum of action corresponds to geodesic equation i.e., $ \delta I = 0$.

Consider $ g_{ij} \dot x^i \dot x^j = \kappa $. Here $\kappa = 0$ corresponds to null propagation of light and $\kappa = 1 $ corresponds to timelike geodesics.

Consider $\theta$ equation of motion of the above action:

\[2r^2\left(\frac{\sqrt{ \left(1-\epsilon ^2 \sin^2 \theta \right)\text{ $\dot \theta $}^2 + \sin^2\theta \text{$\dot \phi $}^2 }}{1-\epsilon ^2 \sin^2\theta}-\frac{ \text{$\epsilon $} sin^2\theta \text{$\dot \phi $}}{1-\epsilon ^2 \sin^2\theta }\right)*\]
\[\left(\frac{2 \epsilon ^2 \sin \theta  \cos\theta  \sqrt{\text{$\dot  \theta $}^2 \left(1-\epsilon ^2 \sin^2\theta\right)+\text{$\dot \phi $}^2 \sin^2 \theta}}{\left(1-\epsilon ^2 \sin^2 \theta \right)^2}-\frac{2 \text{$\dot \phi $} \epsilon ^3  sin^3\theta \cos\theta }{\left(1-\epsilon ^2 \sin^2 \theta \right)^2}+\right.\]
\begin{equation}\left.\frac{(\dot\theta^2\dot)(1-\epsilon^2\sin^2\theta)-2 \text{$\dot \theta $}^2 \epsilon ^2 \sin\theta  \cos \theta + 2 \text{$\dot \phi $}^2 \sin \theta  \cos\theta}{2 \left(1-\epsilon ^2 \sin^2 \theta \right) \sqrt{\text{$\dot \theta $}^2 \left(1-\epsilon ^2 \sin^2 \theta \right)+\text{$\dot \phi $}^2 \sin^2 \theta}}-\frac{2 \text{$\dot \phi $} \text{$\epsilon $}sin\theta cos\theta }{1-\epsilon ^2 \sin^2 \theta }\right) = 0\end{equation}
Assume initially the particle is in $\theta = \frac{\pi}{2}$ plane. Then we get $(r^2\dot \theta^2 \dot) = 0$. Hence, if the particle is initially in equatorial plane i.e.,$ \theta = \frac{\pi}{2}$, then one can always choose a coordinate system such that particle is in the same plane.

So, we put $ \theta = \frac{\pi}{2}$, the metric becomes
\begin{equation} ds^2 = V(r)\dot r^2 - 2\dot v\dot r -r^2\left(\frac{1-\epsilon}{1 - \epsilon^2}\right)^2\dot\phi^2\end{equation}

We consider $r^2\dot \phi = constant = L$ ,$V(r)\dot v - \dot r = -E$.
Here L refers to angular momentum per unit mass and E corresponds to energy of the particle.

Let's consider radial equation of motion:
\begin{equation}g_{ij}\dot x^i\dot x^j = \kappa \end{equation}
\begin{equation}V(r)\dot v^2 - 2\dot v\dot r-r^2\left(\frac{1-\epsilon}{1-\epsilon^2}\right)^2\dot\phi^2 = \kappa \end{equation}
Substituting $\dot v = \frac{-E+\dot r}{V(r)}$,$\dot \phi = \frac{L}{r^2}$ in the above equation,we get
\begin{equation}V(r)\left(\frac{\dot r-E}{V(r)}\right)^2 - 2\dot r\left(\frac{\dot r-E}{V(r)}\right) -r^2\left(\frac{1-\epsilon}{1-\epsilon^2}\right)^2\left(\frac{L}{r^2}\right)^2 = \kappa \end{equation}
\begin{equation}\frac{E^2}{V(r)} - \frac{\dot r^2}{V(r)} - \frac{L^2}{r^2}\left(\frac{1-\epsilon}{1-\epsilon^2}\right)^2 = \kappa\end{equation}
\hfill \break
Let's consider $r = \frac{2M}{u} \implies \dot r = -\frac{2M}{u^2}\dot u = -\frac{2M}{u^2}\frac{du}{d\phi}\frac{l}{r^2} = -\frac{l}{2M}\frac{du}{d\phi}$,$V(r) = 1-u$.
\hfill \break
The equation (20) becomes 
\begin{equation}\left(E^2 - \frac{L^2}{4M^2}\left(\frac{du}{d\phi}\right)^2\right)\frac{1}{1-u} + \frac{L^2}{4M^2}u^2\left(\frac{1-\epsilon}{1-\epsilon^2}\right)^2  = \kappa \end{equation}
\begin{equation}\implies \frac{1}{2}\left(\frac{du}{d\phi}\right)^2 +\left(\left(\frac{u^2}{2} - \frac{u^3}{2}\right)\left(\frac{1-\epsilon}{1-\epsilon^2}\right)^2  - \frac{2m^2u\kappa}{L^2}\right) = \frac{2M}{L^2}\left(E^2 - \kappa\right) \end{equation}

Consider particle motion(of unit mass) to be in 1 dimension with potential U(x).
The total energy is given by the following energy equation:
\begin{equation}Total Energy = \frac{1}{2}\dot x^2 + U(x) \end{equation}
Lets consider motion of photon and massive particle near schwarzschild black hole.
\hfill \break
\textbf{Motion of photon near a Schwarzschild black hole ($\kappa = 0$)} :

The potential is maximum at $u - \frac{3}{2}u^2 = 0 \implies u=\frac{2}{3}$ and $r = \frac{2M}{u} = 3M$.
Hence, there is a circular null geodesic at r=3M.

The potential energy in case of photon is $U(u) = \left(\frac{u^2}{2} - \frac{u^3}{2}\right)\left(\frac{1-\epsilon}{1-\epsilon^2}\right)^2$.
Lets look at comparision of potential energy in this case with schwarzschild case in lorentz metric($ \epsilon = 0 $).
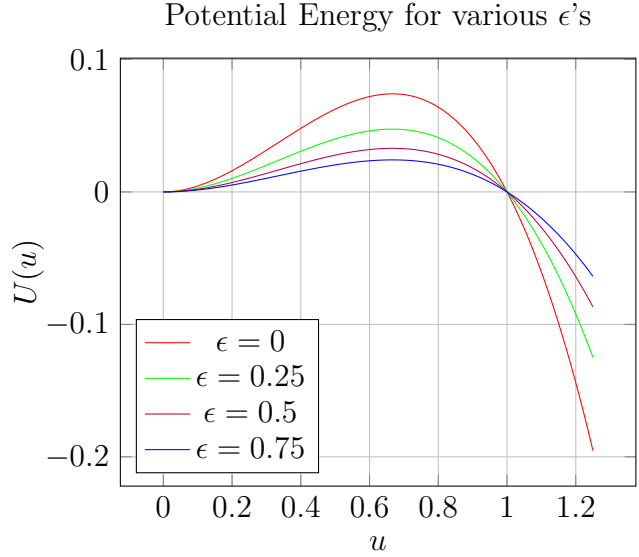
\begin{center}
\begin{tikzpicture}
\begin{axis}[title = Potential Energy for various $\epsilon$'s,domain=0:1.25, samples=100,grid=major,
    restrict y to domain=-4:4,xlabel=$u$,ylabel=$U(u)$, legend pos=south west]
\addplot [color=red]    {(x^2 - x^3)/2 };
\addplot [color=green]  {((x^2 - x^3)/2)*((1-0.25)/(1-0.25^2))^2};
\addplot [color=purple] {((x^2 - x^3)/2)*((1-0.5)/(1-0.5^2))^2}; 
\addplot [color=blue]   {((x^2 - x^3)/2)*((1-0.75)/(1-0.75^2))^2};
\legend{$\epsilon = 0$, $\epsilon = 0.25$, $\epsilon = 0.5$, $\epsilon = 0.75$}
\end{axis}
\end{tikzpicture}
\end{center}
We can see that the maximum potential energy is greater for smaller value of $\epsilon$.
\hfill \break
\textbf{Motion of massive particle near a Schwarzschild black hole ($\kappa = 1$)}:

The potential energy in case of massive particle is $U(u) = \left(\frac{u^2}{2} - \frac{u^3}{2}\right)\left(\frac{1-\epsilon}{1-\epsilon^2}\right)^2  - \frac{2m^2u}{L^2}$.

In this case, we have three possibilities similar to lorentz case:

(i)$\frac{16m^2}{L^2} \left(\frac{1-\epsilon^2}{1-\epsilon}\right)^2 < 1$

\begin{center}
\begin{tikzpicture} 
\begin{axis}[title = Case (i),domain=0:1.25, samples=100,grid=major,
    restrict y to domain=-4:4,xlabel=$u$,ylabel=$U(u)$, legend pos=south west]
\addplot [color=red]    {((x^2 - x^3)/2) - (x/16)};
\addplot [color=green]  {(((x^2 - x^3)/2)-x/16)*((1-0.25)/(1-0.25^2))^2};
\addplot [color=purple] {(((x^2 - x^3)/2)-x/16)*((1-0.5)/(1-0.5^2))^2}; 
\addplot [color=blue]   {(((x^2 - x^3)/2)-x/16)*((1-0.75)/(1-0.75^2))^2};
\legend{$\epsilon = 0$, $\epsilon = 0.25$, $\epsilon = 0.5$, $\epsilon = 0.75$}
\end{axis}
\end{tikzpicture}
\end{center}
In this case, we have turning points for positive values of potential.

(ii)$1 < \frac{16m^2}{L^2} (\frac{1-\epsilon^2}{1-\epsilon})^2 < \frac{4}{3}$

\begin{center}
\begin{tikzpicture} 
\begin{axis}[title = Case (ii),domain=0:1, samples=100,grid=major,
    restrict y to domain=-4:4,xlabel=$u$,ylabel=$U(u)$, legend pos=south west]
\addplot [color=red]    {((x^2 - x^3)/2) - 2*1.10*(x/16)};
\addplot [color=green]  {(((x^2 - x^3)/2)-2*1.10*(x/16))*((1-0.25)/(1-0.25^2))^2};
\addplot [color=purple] {(((x^2 - x^3)/2)-2*1.10*(x/16))*((1-0.5)/(1-0.5^2))^2}; 
\addplot [color=blue]   {(((x^2 - x^3)/2)-2*1.10*(x/16))*((1-0.75)/(1-0.75^2))^2};
\legend{$\epsilon = 0$, $\epsilon = 0.25$, $\epsilon = 0.5$, $\epsilon = 0.75$}
\end{axis}
\end{tikzpicture}
\end{center}
In this case, we have turning points for negative values of potential.

(iii)$\frac{16m^2}{L^2} (\frac{1-\epsilon^2}{1-\epsilon})^2 > 1$

\begin{center}
\begin{tikzpicture}
\begin{axis}[title = Case (iii),domain=0:1, samples=100,grid=major,
    restrict y to domain=-4:4,xlabel=$u$,ylabel=$U(u)$, legend pos=south west]
\addplot [color=red]    {((x^2 - x^3)/2) - 2*2*(x/16)};
\addplot [color=green]  {(((x^2 - x^3)/2)-2*2*(x/16))*((1-0.25)/(1-0.25^2))^2};
\addplot [color=purple] {(((x^2 - x^3)/2)-2*2*(x/16))*((1-0.5)/(1-0.5^2))^2}; 
\addplot [color=blue]   {(((x^2 - x^3)/2)-2*2*(x/16))*((1-0.75)/(1-0.75^2))^2};
\legend{$\epsilon = 0$, $\epsilon = 0.25$, $\epsilon = 0.5$, $\epsilon = 0.75$}
\end{axis}
\end{tikzpicture}
\end{center}
For this case, we have no real turning points.
One could see that as $\frac{M^2}{L^2}$ increases, the peak decreases.
\section{IV.Particle Scattering in Schwarzschild metric in Finsler setting}
Let's consider a schwarzschild black hole in finsler spacetime. Let 'b' be the impact parameter which measures the distance of the particle off the axis of centre of black hole.
\hfill \break
Let's look at scattering of massless and massive particle near the black hole r=2M.
\hfill \break
\textbf{Scattering of photon}:

We can write  $L = r^2 \dot \phi = (r\dot\phi ) r = cb = b $ ( Here c = 1 for photon ).

The total energy in this case = $\frac{2M^2E^2}{L^2}$.
For the photon to be absorbed by black hole,it should be greater than maximum  potential $U_{max} = \frac{2}{27}$ $\implies \frac{ME}{L} > \sqrt{\frac{2}{27}}$.
\begin{equation} L = r^2 \dot \phi = r^2 \frac{d\phi}{dt}\frac{dt}{ds} = Eb \implies \frac{E}{L} = \frac{1}{b}\end{equation}

Hence, if $ M > \frac{b}{\sqrt{27}}$, the photon gets absorbed.

The maximum impact parameter is $b_{max} = 3\sqrt{3}M$.

The cross section for absorption of photon by black hole is $ \sigma_{abs} = \pi b_{max}^2 = 27\pi M^2$
\hfill \break
\textbf{Scattering of massive particle}:

For non-relativistic motion moving with velocity $v_m$, $E = \gamma^{-1} = (1+\frac{1}{v_m^2}+.....) \implies E^2= 1+v_m^2+.....$.

The total energy of massive particle is $\frac{2M^2}{L^2}\left(E^2-\kappa\right) = \frac{2M^2}{L^2}(1+v_m^2+....-1) = \frac{2M^2v_m^2}{L^2}$.

The potential energy in this case is $U(u) = \frac{1}{2}\left(u^2 - u^3\left(\frac{1-\epsilon}{1-\epsilon^2}\right)^2 - \frac{4m^2u}{L^2}\right)$.

The boundary conditions required for the particle to be absorbed by the black hole are :
$U_{max}= 0 $ and $U'(u) = 0 $.
\begin{equation}U_{max}= 0 \implies u^2 - u + \frac{4M^2}{L^2}\left(\frac{1-\epsilon^2}{1-\epsilon}\right)^2 = 0\end{equation}
\begin{equation}U' = 0 \implies 3u^2 - 2u + \frac{4M^2}{L^2}\left(\frac{1-\epsilon^2}{1-\epsilon}\right)^2 = 0\end{equation}

Solving (25) and (26), we get $u = \frac{1}{2}$.

Substitute u in (25) $\implies \frac{16M^2}{L^2} = \left(\frac{1-\epsilon}{1-\epsilon^2}\right)^2 $.
We have $L = v_mb_c \implies b_c^2 = \frac{16M^2}{v_m^2}(\frac{1-\epsilon^2}{1-\epsilon})^2 $.

If $b<b_c$,the massive particle is captured by the black hole.

Capture cross section for massive particle $\sigma_c$ = $\pi b_c^2 = 16\pi\frac{m^2}{v_m^2}(\frac{1-\epsilon^2}{1-\epsilon})^2 $.

As $v_m\rightarrow 0,\sigma_c$ diverges. This is a consequence of attractive nature of gravity.

\section{V.Conclusion}
In this work, we presented finsler version of motion and scattering of massless and massive particles in schwarzschild metric. We showed that maximum potential acquired by a particle depends on the finsler parameter $\epsilon$ and found out to greater for smaller values of $\epsilon$. For the massive particle, there are three different cases and found that as $\frac{M^2}{L^2}$ increases, the peak decreases. In section 4, we calculated capture cross section of the particles by the black hole. All the results agree with riemannian case when $\epsilon = 0$. In the future work, we will investigate the particle motion and scattering in kerr-newman metric in finslerian setting. We would also study in detail about the Tolman-Oppenheimer-Volkoff equation in the context of finsler spacetime.

\section{References}
[1] Shiing-Shen Chern,Finsler geometry is just Riemannian geometry without the quadratic restriction,1996,Volume 43,Notices of the American Mathematical Society.
\hfill \break
[2] C. Pfeifer, Finsler geometric extension of Einstein gravity, MNR Wohlfarth - Physical Review D, 2012 - APS.
\hfill \break
[3] D. Bao, S.S. Chern and Z. Shen, An introduction to Riemann-Finsler geometry, Springer New York 2000.
\hfill \break
[4] I. Bucataru and R. Miron, Finsler-Lagrange geometry, Editura Academiei Romane 2007, Bucharest.
\hfill \break
[5] Z. Chang and X. Li, Phys. Lett. B 668 (2008) 453 [arXiv:0806.2184 [gr-qc]].
\hfill \break
[6] Z. Chang and X. Li, Phys. Lett. B 676 (2009) 173 [arXiv:0901.1023 [gr-qc]].
\hfill \break
[7] S. Weinberg, Gravitation and cosmology, John Wiley 1972.
\hfill \break
[8] Christian Pfeifer, Mattias N. R. Wohlfarth, Finsler spacetimes and gravity, arXiv:1210.2973 [gr-qc].
\hfill \break
[9] G.S. Asanov, Finsler Geometry, Relativity and Gauge Theories, D. Reidel Publishing Company, Dordrecht, Holland, 1985.
\hfill \break
[10] Miguel A. Javaloyes, Miguel Sánchez, Finsler metrics and relativistic spacetimes, arXiv:1311.4770 [math.DG].
\hfill \break
[11] Xin Li and Zhe Chang, An exact solution of vacuum field equation in Finsler spacetime, arXiv:1401.6363 [physics.gen-ph].
\hfill \break
[12] E. Minguzzi, Light cones in Finsler spacetime, arXiv:1403.7060 [math-ph].
\hfill \break
[13] Amir Babak Aazami, Miguel Angel Javaloyes, Penrose's singularity theorem in a Finsler spacetime, arXiv:1410.7595 [math.DG].
\hfill \break
[14] E Minguzzi, Raychaudhuri equation and singularity theorems in
Finsler spacetimes, arXiv:1502.02313 [gr-qc].
\hfill \break
[15] Andrea Fuster, Finsler pp-waves, arXiv:1510.03058 [gr-qc].
\hfill \break
[16] Matias Dahl, An brief introduction to Finsler geometry.
\hfill \break
[17] Rutz, S.F. Gen Relat Gravit (1993) 25:1139.https://doi.org/10.1007/BF00763757.
\hfill \break
[18] Z.K. Silagadze, On the Finslerian extension of the Schwarzschild metric, arXiv:1007.4632 [gr-qc].
\hfill \break
[19] D. Bao, C. Robles, and Z. Shen, J. Diff. Geom. 66, 377 (2004).

\end{document}